\documentclass[%
reprint,
superscriptaddress,
amsmath,amssymb,
aps,
prl,
]{revtex4-1}

\usepackage{siunitx}
\usepackage{graphicx}
\usepackage{dcolumn}
\usepackage{bm}
\usepackage{todonotes}
\usepackage{simplewick}
\usepackage{hyperref}
\usepackage{physics}
\usepackage{verbatim}
\usepackage{float}
\usepackage{array}
\hypersetup{
	colorlinks,
	linkcolor={blue!50!black},
	citecolor={blue!50!black},
	urlcolor={blue!80!black}
}
\graphicspath{{./figures/}}

\usepackage{xspace}
\makeatletter
\DeclareRobustCommand\onedot{\futurelet\@let@token\@onedot}
\def\@onedot{\ifx\@let@token.\else.\null\fi\xspace}

\makeatother

\newcommand{\zbb}{0\nu\beta\beta}
\newcommand{\tbb}{2\nu\beta\beta}
\newcommand{\bb}{\beta\beta}
\newcommand{\E}{E_{3\max}}
\newcommand{\e}{e_{\max}}

\newcommand\Tstrut{\rule{0pt}{2.4ex}}       
\newcommand\Bstrut{\rule[-1.3ex]{0pt}{0pt}} 
\begin{document}
	
	\title{Ab initio neutrinoless double-beta decay matrix elements for $^{48}$Ca, $^{76}$Ge, and $^{82}$Se}%
	
	\author{A. Belley}%
	\affiliation{TRIUMF 4004 Wesbrook Mall, Vancouver BC V6T 2A3, Canada}%
	\affiliation{Department of Physics, McGill University, 3600 Rue University, Montr\'eal, QC H3A 2T8, Canada}%
	\affiliation{Department of Physics \& Astronomy, University of British Columbia, Vancouver, British Columbia V6T 1Z1, Canada}
	
	\author{C. G. Payne}%
	\altaffiliation[Present address: ]{Institut f\"ur Kernphysik and PRISMA$^+$ Cluster of Excellence, Johannes Gutenberg-Universit\"at at Mainz, 55128 Mainz, Germany}
	\affiliation{TRIUMF 4004 Wesbrook Mall, Vancouver BC V6T 2A3, Canada}%
	\affiliation{Department of Physics \& Astronomy, University of British Columbia, Vancouver, British Columbia V6T 1Z1, Canada}
	
	\author{S. R. Stroberg}%
	\affiliation{Department of Physics, University of Washington, Seattle, WA 98195, USA}
	
	\author{T. Miyagi}%
	\affiliation{TRIUMF 4004 Wesbrook Mall, Vancouver BC V6T 2A3, Canada}%
	
	\author{J. D. Holt}%
	\affiliation{TRIUMF 4004 Wesbrook Mall, Vancouver BC V6T 2A3, Canada}%
	\affiliation{Department of Physics, McGill University, 3600 Rue University, Montr\'eal, QC H3A 2T8, Canada}%
	
	\begin{abstract}
		
		We calculate basis-space converged neutrinoless $\beta \beta$ decay nuclear matrix elements for the lightest candidates: $^{48}$Ca, $^{76}$Ge and $^{82}$Se. 
		Starting from initial two- and three-nucleon forces, we apply the ab initio in-medium similarity renormalization group to construct valence-space Hamiltonians and consistently transformed $\beta \beta$-decay operators.
		We find that the tensor component is non-negligible in $^{76}$Ge and $^{82}$Se, and resulting nuclear matrix elements are overall 25-45\% smaller than those obtained from the phenomenological shell model.
		While a final matrix element with uncertainties still requires substantial developments, this work nevertheless opens a path toward a true first-principles calculation of neutrinoless $\beta\beta$ decay in all nuclei relevant for ongoing large-scale searches.
		
	\end{abstract}
	
	\maketitle
	
	Neutrinoless double-beta ($0\nu\beta\beta$) decay is a hypothesized nuclear-weak process in which two neutrons transform into two protons by emitting two electrons~\cite{Avignone2008}. 
	The key feature of this decay is that it produces two leptons (the electrons) without any anti-leptons, thus violating lepton-number conservation. 
	For such a decay to occur, the neutrino must be Majorana, i.e. its own anti-particle~\cite{Furry1938, Schechter1952}. 
	Furthermore, under standard light-neutrino exchange, the rate of the process can be related to the effective neutrino mass $\langle m_{\beta\beta}\rangle$~\cite{Engel2017}:
	\begin{equation}
	[T^{0\nu}_{1/2}]^{-1}=G^{0\nu}|M^{0\nu}|^2 \langle m_{\beta\beta}\rangle ^2,
	\label{eq: Half-Life}
	\end{equation}
	where $G^{0\nu}$ is a phase-space factor whose value is generally agreed upon~\cite{Kotila2012, SUHONEN1998123}.
	Thus an observation could determine the absolute neutrino mass, its Majorana/Dirac character, and most importantly, provide an observation of lepton-number violation, which would have deep implications for the matter-antimatter asymmetry puzzle~\cite{Fukugita1986}.
	
	From Eq.~\ref{eq: Half-Life}, we see that the rate cannot be directly connected to neutrino masses without first having knowledge of the non-observable nuclear matrix element (NME), $M^{0\nu}$, governing the decay. 
	As large-scale searches worldwide will soon enter a ton-scale era probing the inverted neutrino mass hierarchy \cite{Kharusi:2018, Myslik2018, Paton2019, Gando2019, Arnold2010}, a reliable NME with rigorous theoretical uncertainty estimates is imperative not only to pin down $m_{\beta\beta}$, should a discovery be made, but also to interpret evolving experimental lifetime limits in terms of excluded neutrino mass scales. 
	
	Calculations of the NME have proven to be tremendously challenging for nuclear theory, as they require a consistent treatment of nuclear and electroweak forces, as well as an accurate solution of the nuclear many-body problem in heavy systems.
	To date, almost all calculations of $\zbb$ decay have been based on nuclear models, but since no $\zbb$-decay data exist to constrain these models, unsurprisingly a spread in results (up to factors of three) has emerged~\cite{Engel2017,MENENDEZ2009139,Simkovic2013,Senkov2014,Vaquero2014,Barea2015}.
	This spread is \emph{not} a true uncertainty, however, as all models are known to neglect essential physics. 
	Since experimental expectations for material and timescale requirements are based on the currently available spread, they may need to be reevaluated should improved values lie well outside the existing range. 
	Therefore it is critical to have next-generation NMEs for the most prominent experimental candidates -- $^{76}$Ge, $^{130}$Te and $^{136}$Xe -- to guide next-generation searches.
	
	Chiral effective field theory (EFT)~\cite{Epelbaum2009, Machleidt2011} in principle provides a prescription for the consistent treatment of nuclear forces and electroweak currents relevant for $\zbb$ decay~\cite{Menendez2011,Cirigliano2018,Cirigliano2018a}.
	While first calculations have been carried out in the lightest nuclei~\cite{Cirigliano2019,Wang2019}, the only calculations of experimental $\zbb$-decay candidates from chiral forces have been in a perturbative shell-model effective-interaction framework~\cite{Holt2013,Kwiatkowski2014,Jiao2017,Coraggio2020}. 
	While results were encouraging, order-by-order convergence was unclear. 
	With the advent of nonperturbative theories capable of reaching at least $A=100$~\cite{Hagen2014,Hergert2016,Morris2018,Stroberg2019}, the primary bottleneck has been the computational resources needed to obtain converged results and the treatment of deformed systems. 
	With ongoing advances in the field, the first ab initio calculations of $\zbb$-decay are within reach, and indeed very recently NMEs were reported for $^{48}$Ca in the in-medium generator coordinate method (IM-GCM)~\cite{Yao2020}.
	
	\begin{figure*}[t]
		\includegraphics[width=\textwidth]{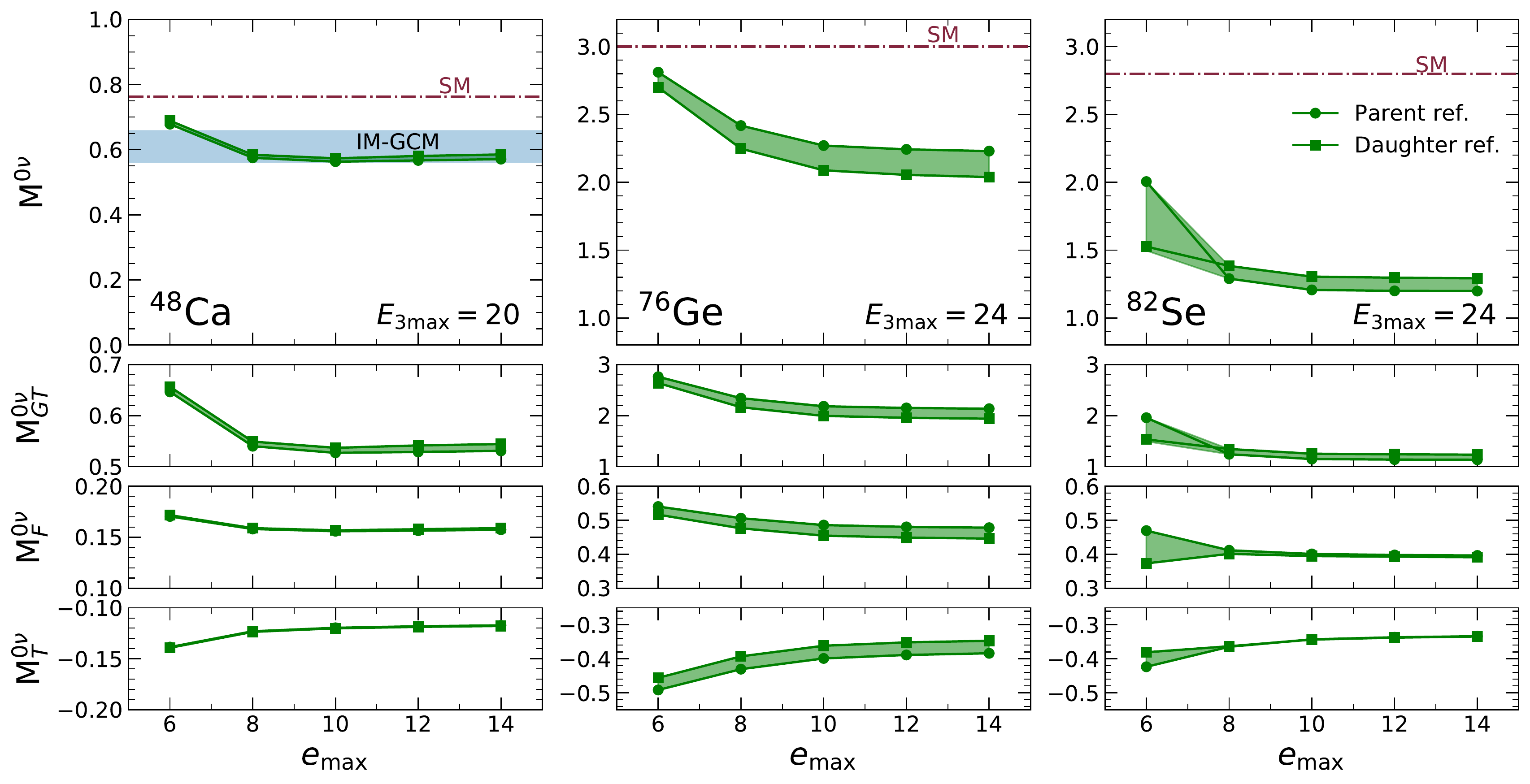}
		\caption{NMEs for the $\zbb$-decay of $^{48}$Ca, $^{76}$Ge, and $^{82}$Se as a function of $\e$, at fixed $\E$. The bands represent the uncertainty from the choice of ENO reference. We also show the convergence of the GT, F  (with factor  of $-(\frac{g_v}{g_a})^2$ included), and T operators separately. In addition we compare to phenomenological shell-model (SM) results quoted in Ref.~\cite{Engel2017} 
			for each decay and to complementary ab initio IM-GCM values \cite{Yao2020} (blue band) in $^{48}$Ca, which agree within uncertainties.}
		\label{fig:NMES_total}
	\end{figure*}
	
	In this Letter we extend ab initio calculations of the NMEs to the three lightest $\bb$-decay nuclei $^{48}$Ca, $^{76}$Ge and $^{82}$Se using the valence-space in-medium similarity renormalization group (VS-IMSRG) \cite{Tsukiyama2012,Hergert2016, Bogner2014, Stroberg2016,Stroberg2017,Stroberg2019}.
	We first demonstrate convergence in terms of the single-particle basis size and truncations imposed on three-nucleon (3N) forces. 
	In contrast to phenomenology, we find that the tensor operator is non-negligible for $^{76}$Ge and $^{82}$Se and is approximately the same magnitude as the Fermi term.
	As seen in Fig.~\ref{fig:NMES_total}, the NMEs are smaller than standard shell-model calculations by approximately 25\% in $^{48}$Ca, 30\% in $^{76}$Ge, and 45\% in $^{82}$Se, but in remarkably good agreement with IM-GCM and coupled cluster theory~\cite{Novario2020,Yao2020} in $^{48}$Ca when starting from the same input forces.
	
	The $\zbb$-decay operator under standard light neutrino exchange is given by the sum of the allowed Gamow-Teller (GT), Fermi (F), and tensor (T) transitions~\cite{MENENDEZ2009139}:
	\begin{equation}
	M^{0\nu} = M^{0\nu}_{GT} - \Big(\frac{g_V}{g_A}\Big)^2 M^{0\nu}_{F} + M^{0\nu}_{T}
	\end{equation}
	where $g_V$=1 and $g_A$=1.27 are the unquenched vector and axial coupling constants, respectively. 
	Explicit NME expressions and details can be found in Refs.~\cite{Engel2017,Doi1985, Tomoda1991,Horoi2010, Cirigliano2018a}.
	To avoid explicit sums over intermediate states, we use the standard closure approximation, with ``closure energy" $\Bar{E} \approx E_k - (E_i + E_f)/2$.
	Corrections to the closure approximation are of order $\Bar{E}/q\sim 10\%$~\cite{Senkov2013}, with momentum exchange $q\sim 1$~fm, and weakly dependent on the choice of $\Bar{E}$~\cite{Horoi2010}.
	Within the framework of chiral EFT, these corrections appear at sub-leading order~\cite{Cirigliano2018a}.
	To facilitate benchmarking with previous calculations, we used a value $\Bar{E}=7.72$ MeV for $^{48}$Ca and $\Bar{E}=9.41$~MeV for the heavier isotopes~\cite{Haxton1984,Tomoda1991}.
	We also use dipole form factors with cutoffs $\Lambda_{V}=850$~MeV and $\Lambda_{A}=1086$~MeV~\cite{Simkovic2009}, and multiply the NMEs by the nuclear radius $R=1.2 A^{1/3}$~fm to make them dimensionless~\cite{Engel2017}.
	The necessity of a leading-order short-range contact term has recently been discovered~\cite{Cirigliano2018,Cirigliano2018a} and has yet to be included in any calculations of experimental candidates. 
	Preliminary assessments of its importance show the effect can be as large as the full matrix element, 
	so studies of subleading two-body currents will not be relevant until there is a firm handle on this value. 
	
	We calculate NMEs from two-nucleon (NN) plus 3N forces from chiral EFT. 
	In particular we use 1.8/2.0(EM) from a family of  Hamiltonians~\cite{Hebeler2011, Simonis2016,Simonis2017}, where 3N couplings are constrained by the binding energy of $^{3}$H and the charge radius of $^{4}$He. 
	This interaction globally reproduces ground-state energies to the tin isotopes, including the nuclear driplines in the light- and medium-mass regions, albeit while giving radii that are systematically too small compared to experiment \cite{Morris2018,Simonis2017,Holt2019}. 
	
	We begin in a harmonic-oscillator (HO) basis with $\hbar\omega=16$ MeV and $e= 2n+l \le e_{\mathrm{max}}$ with a cut of $e_1+e_2+e_3 \le E_{3\mathrm{max}}$ on 3N matrix elements. 
	We transform the Hamiltonian and $\zbb$-decay operator to the Hartree-Fock (HF) basis, accounting for 3N forces between valence nucleons via ensemble normal-ordering (ENO)~\cite{Stroberg2017} at the two-body level (NO2B). 
	Previous $\E$ limitations were 16 or 18, but we are now able to routinely calculate with $\E=24$ or higher~\cite{Miyagi_inprep}, putting heavy nuclei within reach.
	We use the Magnus formulation of the IMSRG \cite{Morris2015,Hergert2016} to derive an approximate unitary transformation to decouple a valence-space Hamiltonian~\cite{Tsukiyama2012,Bogner2014,Stroberg2019}, and consistently transformed operators~\cite{Parzuchowski2017}. 
	We use the IMSRG(2) approximation where all operators are truncated at the NO2B level.
	We take the standard $pf$-shell valence space for $^{48}$Ca and the $p_{1/2}$, $p_{3/2}$, $f_{5/2}$, $g_{9/2}$ proton and neutron orbits outside a $^{56}$Ni core for $^{76}$Ge and $^{82}$Se.
	The valence-space diagonalization is done using the KSHELL shell-model code~\cite{Shimizu2019}.
	
	\begin{figure}[t]
		\centering
		\includegraphics[width=1.0\columnwidth]{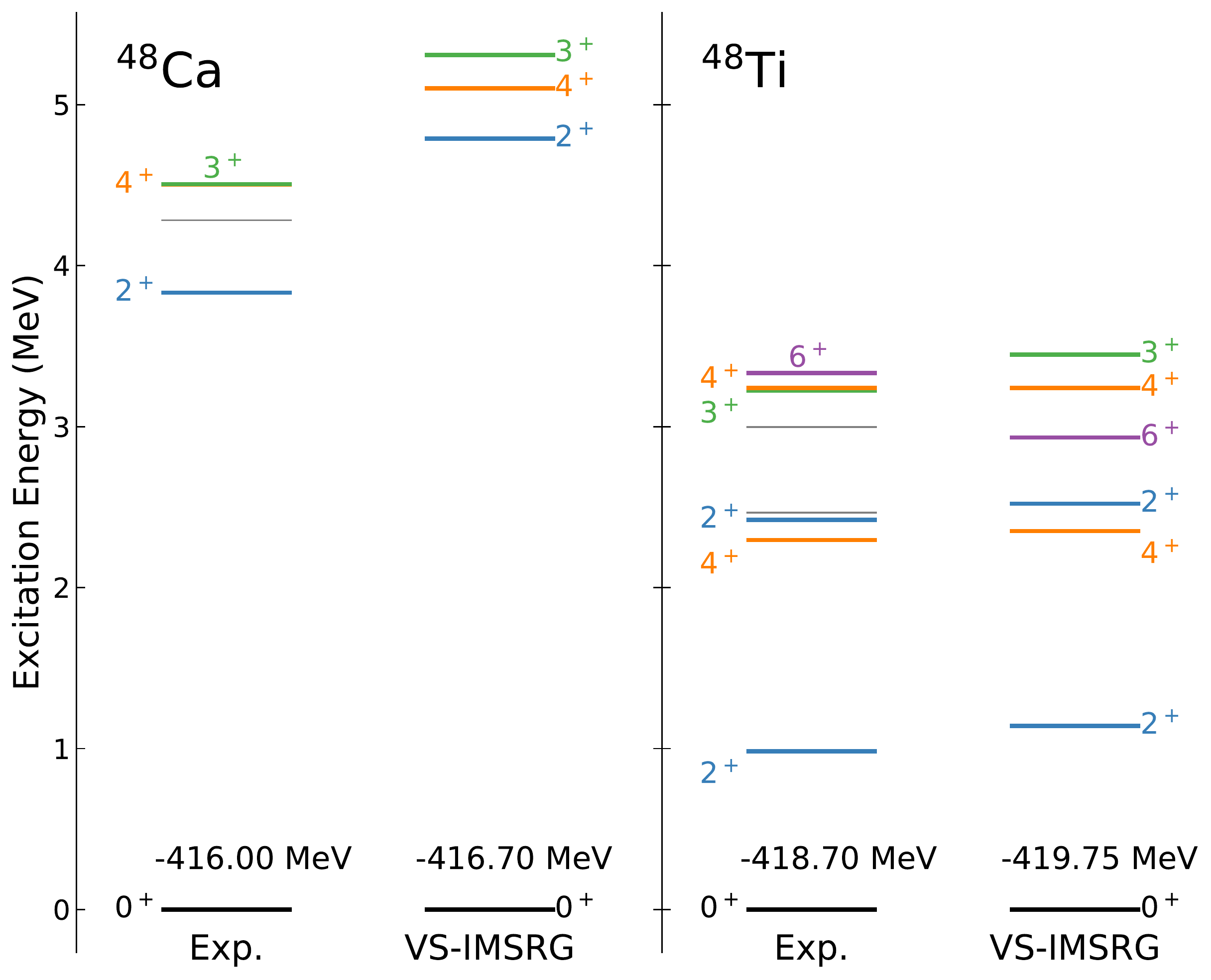}\\
		\vspace{0.5cm}
		\includegraphics[width=1.0\columnwidth]{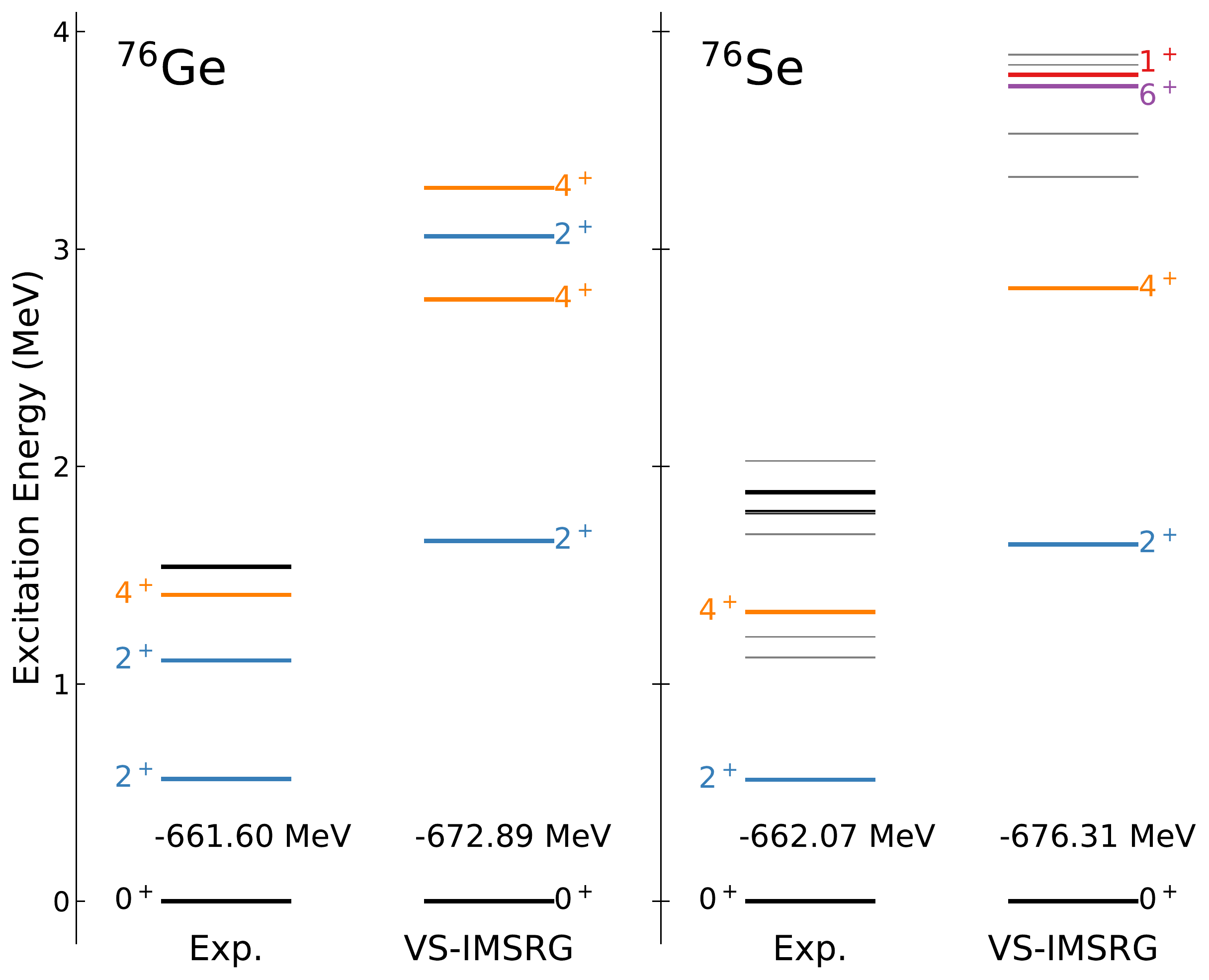}
		\caption{Excitation spectra of  $^{48}$Ca/Ti and $^{76}$Ge/Se from the VS-IMSRG compared to experimental values \cite{Wang2017, Wang2012}. Certain states have been highlighted to help guide the comparison.}
		\label{fig:Spectrum}
	\end{figure}
	
	Before addressing $\zbb$ decay, we must first validate and benchmark in as many relevant electroweak processes as possible.
	For the longstanding puzzle of $g_A$ quenching in nuclei, which still persists in discussions of $\zbb$ decay, we have recently shown that for a wide range of nuclei, when two-body currents consistent with input Hamiltonians are included in ab initio calculations, experimental GT transitions are largely reproduced with an unmodified $g_A$~\cite{Gysbers2019}. 
	We have also calculated the $\tbb$ decay of $^{48}$Ca and find a preliminary value of 0.025 for the effective NME, which is modestly smaller than the experimental value of 0.035 $\pm$ 0.003 obtained in Ref.~\cite{Barabash_2020}. 
	We anticipate that there will be contrasting effects from currently neglected physics: two-body currents will likely lower the NME as in GT quenching~\cite{Gysbers2019}, while higher-order many-body effects, i.e., from IMSRG(3), have been shown to increase the NME in coupled-cluster theory~\cite{Novario2020}. 
	An in-depth analysis and benchmark is currently in preparation~\cite{2v_inprep}.  
	Finally we have benchmarked fictitious $\zbb$-decay rates in light nuclei for selected systems from $A=6$ to $A=22$~\cite{Yao20light}. 
	Comparing with results from no-core shell model, coupled-cluster theory and IM-GCM, we find discrepancies are typically less than 10\%, with somewhat larger deviations found in $^{8}$He and $^{14}$C.
	Therefore, it appears that the physics expected to be relevant for $\zbb$ decay is largely under control to justify first ab initio explorations in heavier experimental candidates. 
	
	In Fig.~\ref{fig:Spectrum} we show the excitation spectra for both parent and daughter nuclei compared to the experimental values for the $^{48}$Ca and $^{76}$Ge transitions (the spectrum of $^{82}$Se is similar to that of $^{76}$Ge). 
	We see that for the $A=48$ cases, the computed spectra are in good agreement with experiment, similar to the IM-GCM~\cite{Yao2020}.
	Only the first excited state in $^{48}$Ca is several hundred keV too high, but the IMSRG(2) approximation is known to produce too high first excited states in doubly magic systems~\cite{Simonis2017,Taniuchi2019}.
	Otherwise the spectrum of $^{48}$Ti is very well reproduced, implying the collective nature of the nucleus is adequately captured, similar to observations in the $sd$ shell~\cite{Stroberg2016}.
	For the heavier cases, however, the computed spectra are too spread compared to experiment, likely due to missing collectivity.
	Further benchmarks are underway, but from IM-GCM studies, only a weak correlation was seen between NMEs and $(E2)$ strength~\cite{Yao2020}.
	For the $A=48$ systems, the calculated ground-state energies agree with experiment to better than 1\% and the $Q$-value to 300keV, while for $A=76, 82$ the ground states agree to 2\% and $Q$-values to 3~MeV and 4~MeV, respectively.
	
	\begin{figure}[t]
		\centering
		\includegraphics[width=1.0\columnwidth]{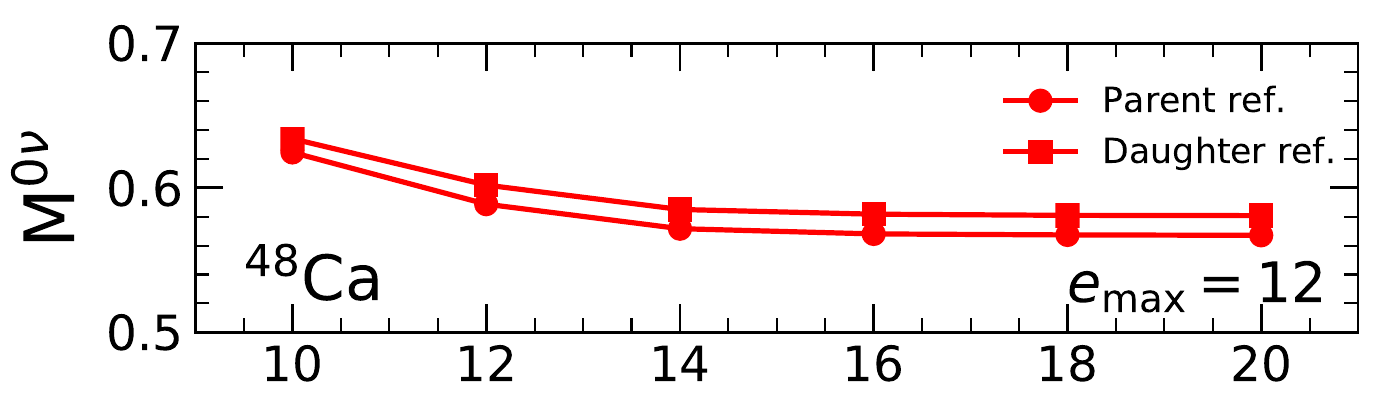}
		\includegraphics[width=1.0\columnwidth]{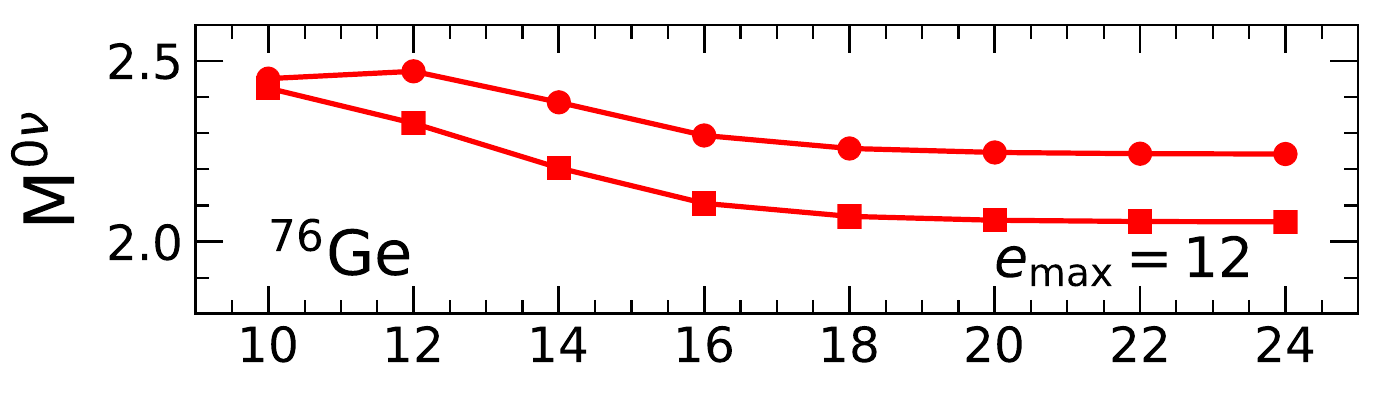}
		\includegraphics[width=1.0\columnwidth]{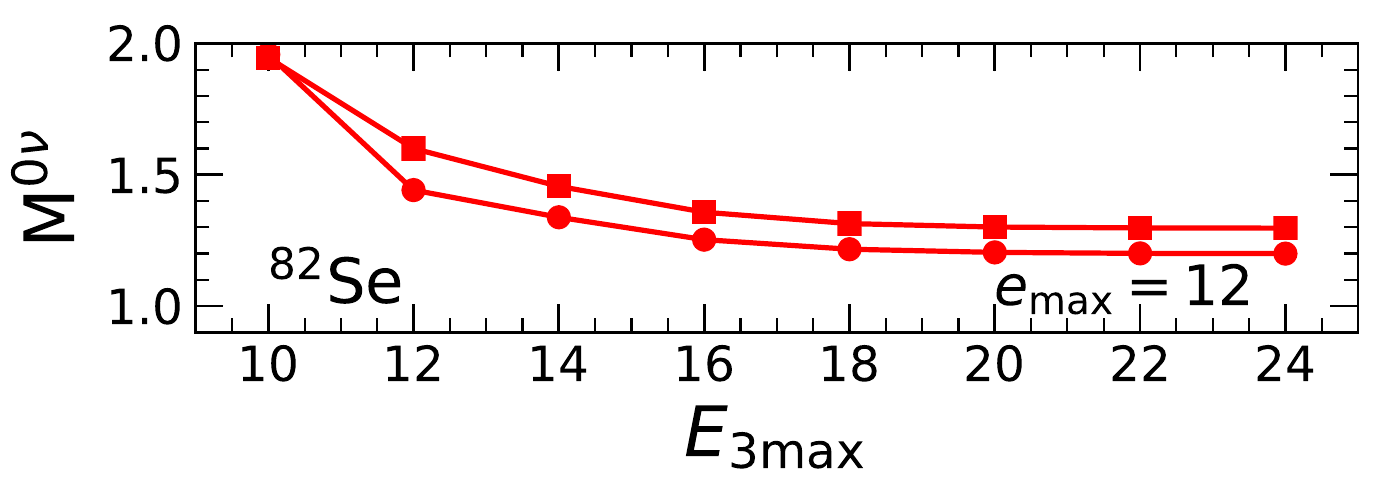}
		\caption{Convergence of the NMEs as we vary the size of the 3-body storage truncation $\E$  at fixed $\e$. As we see, convergence is obtained at $\E=20$ in $^{48}$Ca and $\E=24$ for the heavier isotopes, validating the choices in Fig.~\ref{fig:NMES_total}.}
		\label{fig:E3max_convergence}
	\end{figure}

	\begin{table*}[t]
		\begin{tabular*}{2\columnwidth}{l@{\extracolsep{\fill}}S[table-format= +1.2(1)]S[table-format= +1.2(1)]S[table-format= +1.2(1)]S[table-format= +1.2(1)]S[table-format= +1.2(1)]S[table-format= +1.2(1)]S[table-format= +1.2(1)]S[table-format= +1.2(1)]S[table-format= +1.2(1)]}
			\hline\hline
			\Tstrut & \multicolumn{3}{>{\centering}p{0.49\columnwidth}}{$^{48}$Ca} & \multicolumn{3}{>{\centering}p{0.5\columnwidth}}{$^{76}$Ge} & \multicolumn{3}{>{\centering}p{0.5\columnwidth}}{$^{82}$Se} \\ 
			& HO & HF & IMSRG & HO & HF & IMSRG & HO  & HF & IMSRG\\ 
			\cline{2-4} \cline{5-7} \cline{8-10} 
			GT & 0.51(1)     & 0.46(1)     & 0.54(1) & 4.2(2)    & 3.5(2)    & 2.04(10)  & 3.39(1)   & 2.76(1)   & 1.19(5)     \\
			F & 0.13(1)     & 0.13(1)     & 0.16(1)    & 0.47(1)    & 0.42(1)    & 0.46(2)   & 0.39(1)   & 0.35(1)   & 0.39(1)    \\
			T  & -0.07(1)    & -0.08(1)    & -0.12(1)   & -0.04(1)   & -0.02(1)   & -0.37(2)  & -0.04(1)  & -0.02(1)  & -0.33(1)  \\
			Total\Bstrut & 0.57(1)     & 0.51(1)     & 0.58(1)    & 4.6(2)    & 3.9(2)    & 2.14(9)   & 3.77(1)   & 3.09(1)   & 1.24(5)     \\
			\hline\hline
		\end{tabular*}
		
		\caption{Decomposition of the NMEs for $^{48}$Ca, $^{76}$Ge, $^{82}$Se into their Gamow-Teller (GT), Fermi (F) and tensor (T) part at $e_{\mathrm{max}}$ = 14 and $E_{3\mathrm{max}} = 20$ for $^{48}$Ca and $E_{3\mathrm{max}} = 24$ for $^{76}$Ge and $^{82}$Se. For the Fermi part, the factor of $-(\frac{g_v}{g_a})^2$ as been included. We present the values for the operator in the HO and HF bases with the IMSRG-evolved wavefunctions as well as the fully evolved IMSRG results (IMSRG). The uncertainty represents the range due to the choice of reference state.}
		
		\label{tab:NME}
	\end{table*}
	
	Turning to our $\zbb$-decay results, Fig.~\ref{fig:NMES_total} shows the computed NMEs of $^{48}$Ca, $^{76}$Ge and $^{82}$Se. 
	Here we see clear convergence by $\e=14$ for the total matrix element as well as the three components of the decay. 
	Since the ENO procedure takes a specific nucleus as the reference, we also examine this reference-state dependence. 
	While it is negligible in $^{48}$Ca, there can be changes of up to 10\% in the heavier nuclei.
	We also note that ordering of HF single-particle levels can change with increasing $\e$, which changes the occupations taken for the ENO procedure, as observed between $e_{\mathrm{max}}=6-8$ for $^{82}$Se. 
	The reference-state dependence is expected to be reduced with the introduction of three-body operators in the VS-IMSRG(3).
	In Fig.~\ref{fig:E3max_convergence} we show convergence with $\E$. 
	While $^{48}$Ca is well converged to better than 0.01 in the overall matrix element by $\E  = 16$, perhaps somewhat unexpectedly $\E=20$ is necessary to achieve the same level of convergence in both $^{76}$Ge and $^{82}$Se.
	
	
	Taking a more detailed look at the NME values, we refer to Table~\ref{tab:NME}, where we break down the GT, F, and T components for the unevolved $\zbb$-decay operator in both the HO and HF bases with (albeit inconsistent) VS-IMSRG wavefunctions, as well as the final IMSRG-transformed operator consistent with the wavefunctions. 
	In $^{48}$Ca, we find that the tensor part of the NME, which has been largely neglected in the past~\cite{Kortelainen2007} or found to be negligible by phenomenological methods \cite{Senkov2016-2}, accounts for 20\% of the total matrix element, a modest increase from its contribution in the HO and HF bases. 
	
	For $^{76}$Ge and $^{82}$Se we observe very similar patterns. 
	In previous phenomenological studies, the tensor component is taken or shown to be negligible, which is what we find for HO and HF pictures. 
	However, the IMSRG induces a significant tensor component, reducing the value of the total NME by 15-20\%. 
	While the F part is largely unaffected in all cases, there is also a significant reduction in the GT component.
	The final NME is reduced by a factor of more than two, and we see that correlations do not always have a consistent effect in different systems.
	Since operator reduced matrix elements are combined with two-body transition densities (TBTDs), it can be difficult to trace the origin of these changes.
	For GT transitions, the HF and IMSRG transformations reduce the norm of the valence-space operator matrix elements to approximately 80\% then 25\% and 65\% of their original HO values, respectively.
	The F elements change similarly, but the dominant operator matrix elements are suppressed by small TBTDs, and we see little overall change. 
	In the tensor part, a fine-tuned cancellation that arises in the HO and HF pictures is spoiled in the IMSRG due to the reduction of one matrix element.
	
	The fact that the evolution of the two-body operator leads to such a significant change in the final NMEs highlights the need to investigate the effects of three-body operators.
	We expect the contribution of $n$-body operators to diminish with increasing $n$, but since there is no one-body term to compare,
	estimating the magnitude of three-body terms is crucial to ensure that the two-body term is dominant. 
	Therefore, before claiming final results for the NME, we must first assess the importance of three-body terms in IMSRG(3).
	
	Comparing to the phenomenological shell model quoted in Ref.~\cite{Engel2017}, we see an overall reduction: 25\% in $^{48}$Ca, 30\% in $^{76}$Ge, and 45\% in $^{82}$Se, making the NMEs presented here among the smallest ever reported for these three nuclei.
	This appears to be an emerging picture from complementary ab initio theories. 
	Starting from the same 1.8/2.0(EM) interaction, and employing the same IMSRG(2) approximation, our NME for $^{48}$Ca is completely consistent with the IM-GCM findings in Ref.~\cite{Yao2020} (seen in Fig.~\ref{fig:NMES_total}), as well as preliminary results from coupled-cluster theory~\cite{Novario2020}. 
	Furthermore the NME for $^{76}$Ge again appears to be consistent with preliminary IM-GCM results at the same level of many-body approximation~\cite{Yao_priv}. 
	
	In conclusion, we have computed $\zbb$-decay NMEs for $^{48}$Ca, $^{76}$Ge and $^{82}$Se, finding convergence by $\e=14$ and $\E=20$ with overall smaller values compared to the phenomenological shell model by 25-45\%. While $^{48}$Ca is not a primary experimental candidate, its relatively light mass and doubly magic nature make it a valuable benchmark for various ab initio theories going forward.  
	With the VS-IMSRG advances presented here, we have now provided ab-initio NME computations for the first of three major players in experimental searches: $^{76}$Ge.
	With capabilities to perform calculations at high $\E$, we are already poised to provide NMEs for $^{130}$Te and $^{136}$Xe at the same level as in this work.
	
	Significant work remains to assess all relevant sources of theoretical uncertainty before any claims to a final NME can be made.
	Since the current Hamiltonian is given at the N$^3$LO NN and N$^2$LO 3N levels, we must quantify uncertainties from neglected higher orders. 
	A first step would be to examine the dependence of the NMEs on a wide range of input NN+3N forces. 
	While informative, a more rigorous approach would be an order-by-order analysis of both the Hamiltonian and consistently derived currents with uncertainties, once a reliable calculation of the leading order contact~\cite{Cirigliano2018,Cirigliano:2020} is achieved.
	We have implemented consistent free-space SRG evolution of the $\zbb$-decay operator and are currently investigating the possible importance of induced three-body terms~\cite{Miyagi_inprep}. 
	This will also allow the future study of non-standard neutrino exchanges, which are typically short range in nature.
	Finally development of the IMSRG(3) is underway, which will provide a handle on many-body uncertainties by allowing full treatment of 3N forces beyond the NO2B approximation.
	While difficult to estimate, we would naively expect behavior similar to that observed in coupled cluster calculations where going from doubles to approximate triples corrections increases the NME in $^{48}$Ca by a modest 0.05.
	Only once this program has been accomplished and complementary many-body approaches can each produce independent predictions with uncertainty estimates, can the field give a firm statement on NMEs for experimental $\zbb$ searches.
	
	\begin{acknowledgments}
		
		We thank J. Engel, G. Hagen, H. Hergert, M. Horoi, B. Hu, J. Men\'endez, P. Navr\'atil, S. Novario, T. Papenbrock, N. Shimizu, and J. M. Yao for enlightening discussions and extensive benchmarking, and K. Hebeler for providing momentum-space inputs for generation of the 3N forces used in this work.
		The IMSRG code used in this work makes use of the Armadillo \texttt{C++} library \cite{Sanderson2016}.
		TRIUMF receives funding via a contribution through the National Research Council of Canada. 
		This  work was further supported by NSERC, the Arthur B. McDonald Canadian Astroparticle Physics Research Institute, the Canadian Institute for Nuclear Physics, and the US Department of Energy (DOE) under contract DE-FG02-97ER41014.
		Computations were performed with an allocation of computing resources on Cedar at WestGrid and Compute Canada, and on the Oak Cluster at TRIUMF managed by the University of British Columbia department of Advanced Research Computing (ARC).
		
	\end{acknowledgments}
	
	\bibliographystyle{apsrev4-1}
	\bibliography{library}{}
\end{document}